# Tailoring Charge Donor-Acceptor Interaction in CsPbBr$_3$ Perovskite Nanocrystals through Ligand Exchange

Syed Abdul Basit Shah[1], Sushant Ghimire[2,3*], Rostyslav Lesyuk[2,4], Maria Vittoria Diamanti[5], Vanni Lughi[1] and Christian Klinke[2,6,7*]

[1]Department of Engineering and Architecture, Università degli studi di Trieste, Trieste, Italy;

[2]Institute of Physics, University of Rostock, 18059 Rostock, Germany;

[3]Chair for Photonics and Optoelectronics, Nano-Institute Munich and Department of Physics, Ludwig-Maximilians-Universität (LMU), 80539 Munich, Germany;

[4]Pidstryhach Institute for Applied Problems of Mechanics and Mathematics of NAS of Ukraine, Lviv, Ukraine;

[5]Department of Chemistry, Materials and Chemical Engineering "Giulio Natta" Politecnico di Milano, Italy;

[6]Department of Chemistry, Swansea University, Swansea, United Kingdom;

[7]Department "Life, Light & Matter", University of Rostock, 18059 Rostock, Germany.

*Corresponding Authors: christian.klinke@uni-rostock.de; sushant.ghimire@lmu.de

## ABSTRACT

The surface ligands in colloidal metal halide perovskites influence not only their intrinsic optoelectronic properties but also their interaction with other materials and molecules. We explore donor-acceptor interactions of $CsPbBr_3$ perovskite nanocrystals with $TiO_2$ nanoparticles and nanotubes by replacing long-chain oleylamine ligands with short-chain butylamines. Through post-synthesis ligand exchange, we functionalize the nanocrystals with butylamine ligands while maintaining their intrinsic properties. In solution, butylamine-capped nanocrystals exhibit reduced photoluminescence intensity with increasing $TiO_2$ concentration but without any change in photoluminescence lifetime. Intriguingly, the Stern-Volmer plot depicts different slopes at low and high $TiO_2$ concentrations, suggesting mixed static and sphere-of-action quenching interactions. Oleylamine-capped nanocrystals in solution, on the other hand, show no interaction with $TiO_2$, as indicated by consistent photoluminescence intensities and lifetimes before and after $TiO_2$ addition. In films, both types exhibit decreased photoluminescence lifetime with $TiO_2$, indicating enhanced donor-acceptor interactions, which are discussed in terms of trap state modification and electron transfer. $TiO_2$ nanotubes enhance nonradiative recombination more in butylamine-capped $CsPbBr_3$ perovskite nanocrystals, emphasizing the role of ligand chain length.

**KEYWORDS.** Metal halide perovskites, ligand exchange, donor-acceptor interaction, charge transfer, Mixed photoluminescence quenching, Quenching by Sphere-of-action

**TOC GRAPHIC**

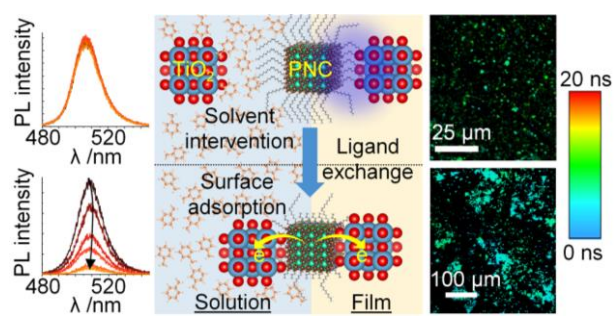

**INTRODUCTION**

Three-dimensional (3D) ABX$_3$ perovskites are outstanding semiconductors due to their excellent optoelectronic properties and diverse applications, including photovoltaics, LEDs, photodetectors, lasers, and photocatalysis.[1] Here, A is a monovalent cation such as Cs$^+$, methylammonium (MA), or formamidinium (FA) occupying a cavity formed by the cornersharing (BX$_6$)$^{4-}$ octahedra consisting of a divalent metal ion B$^{2+}$ (B=Pb, Sn, Ge) and halide ions X$^-$ (X=Cl, Br, I). Among the various compositions of metal halide perovskites, all-inorganic perovskite nanocrystals (PNCs) have gained the spotlight on account of their notable stability, nearly unity photoluminescence (PL) quantum yield (QY), and compatibility with optoelectronic devices.[2] Manipulating the surface of these PNCs through ligand engineering is necessary to optimize their functionality and expand their applications.

Surface ligands play a crucial role in regulating not only the shape and the size of ABX$_3$ PNCs[3] but also their optoelectronic properties, stability, and surface chemistry.[2,4] Surface passivation of PNCs by ligands stabilizes the intrinsic defects as well as prevents molecules such as water and oxygen from reaching their surfaces. This on one hand suppresses the nonradiative decay channels for excitons and hence improves the PL QY and charge carrier dynamics of PNCs, and on the other hand, enhances their intrinsic as well as their environmental stability. While oleic acid and oleylamine are the conventionally used combination of Lewis acid-Lewis base-based binary ligands for synthesizing shape- and size-controlled, stable, and brilliantly luminescent PNCs and quantum dots (QDs),[5] other ligands including phosphine/phosphine oxide,[6] sulfonate,[7] quarternary ammonium,[8] and zwitterionic or bidentate ligands[2c,9] are also used either in combination with the above ones or independently to realize high-quality nanocrystals. In all cases, the role of surface ligands in stabilizing and improving the properties of ABX$_3$ PNCs is crucial.

Besides shape and size control, defect passivation, and stability, the role of surface ligands is important when considering the photophysical interaction of PNCs among themselves and or with other semiconductor materials, metals, and molecules that may result in a charge or energy transfer.[10] The presence of bulky surface ligands may act as an insulating layer for charge transfer between PNCs and other charge-accepting/donating species. This greatly influences the efficiency of devices where $ABX_3$ PNCs are used as light-absorbing or light-emitting layers.[4,11] Also, such ligands may limit the use of PNCs in photocatalysis where exciton dissociation and charge separation are imperative to conduct heterogeneous photoredox reactions.[10g] On the other hand, the removal of excess ligands from the PNC surface may improve the charge transfer process but compromise the stability and optoelectronic properties.[12] Therefore, a balance is required for controlled charge transfer processes in PNCs with minimal influence on their stability and properties.

The use of certain aromatic or conjugated ligands in the synthesis of metal halide perovskites can promote charge delocalization.[10a,d,e,h,i,13] Vickers et al. demonstrated that the film of $MAPbBr_3$ QDs synthesized using benzylamine and benzoic acid ligands showed superior electrical conductivity, extended carrier lifetime, and effective charge transfer to a fluorine-doped tin oxide (FTO) glass substrate, as opposed to QDs with non-conductive ligands.[10a] Also, passivation of the $ABX_3$ PNC surface with short-chain ligands during the synthesis can facilitate better electronic coupling among PNCs or with other charge donor/acceptor species, enhancing the efficiency of charge transfer.[10c,d,12a] Kumar et al. showed an increase in current density and luminance and a decrease in turn-on voltage of an $FA_{0.5}MA_{0.5}PbBr_3$-based LED while decreasing the chain length of the aliphatic amine ligand from sixteen to six carbon.[10c] They attributed the effect of ligand chain length on the enhanced electroluminescence characteristics of $FA_{0.5}MA_{0.5}PbBr_3$ PNC film to an accelerated rate of charge transfer between nanocrystals, which further enhanced charge injection into the PNC layer within the LED device.

Controlling the morphology of 3D ABX$_3$ PNCs through the use of conjugated/aromatic ligands or short-chain ligands during synthesis can be a complex task. Short-chain ligands, for example, may lead to the formation of quasi-2D perovskite nanoplatelets or a mixture of nanoplatelets and nanocubes.[3,14] Similarly, employing short-chain, conjugated, or aromatic ligands might also result in the development of layered 2D Ruddlesden-Popper structures instead of the desired 3D perovskite nanocubes.[15] These low-dimensional metal halide structures hold promise for optoelectronic properties, albeit differing from those of 3D ABX$_3$ PNCs.[16] This disparity makes it challenging to directly compare the charge transfer process in long-chain and short-chain ligand-capped PNCs and metal halide nanostructures. Such a comparison is crucial for a deeper understanding of photophysics in charge donor-acceptor systems based on 3D ABX$_3$ PNCs and is essential for their effective utilization in light emitting and light harvesting applications. Post-synthesis ligand exchange, on the other hand, could be an effective strategy for manipulating the surface of ABX$_3$ PNCs for efficient charge transfer while preserving their morphology and properties.[4,17]

Previous studies discussed different ligand binding modes on the ABX$_3$ PNC surface, suggesting that the organic acid and organic amine ligands are weakly attached to the PNCs and exhibit highly dynamic interactions with the PNC surface.[4,18] This allows for easy ligand exchange, with the native ligands on the surface of PNCs being replaced with the new desired ones. As a result, charge transfer from PNCs to other semiconductors or molecules can be enhanced by precisely tuning the surface ligand length through post-synthesis ligand exchange.[10d,f,19] Biswas et al. successfully demonstrated the post-synthesis exchange of oleic acid in CsPbBr$_3$ PNCs with short-chain benzoic acid and ascorbic acid ligands.[10d] While both the ligand-exchanged samples showed improved stability and enhanced optoelectronic properties, the benzoic acid ligand-capped $CsPbBr_3$ PNCs exhibited better charge transfer rates to the acceptor molecules such as *para*-benzoquinone (electron acceptor) and phenothiazine (hole acceptor) due to the conjugated and short ligand chain. Similarly, Dey et al. showed improved electronic coupling, and therefore enhanced charge transfer across the $CsPbBr_3$ PNC-CdSe nanoplatelet heterojunction which is cross-linked via *para*-aminobenzoic acid or glycine ligands post-synthetically through ligand exchange when compared with their mixture with native long-chain oleic acid and oleylamine ligands.[19]

In this study, we highlight the enhanced interaction between $CsPbBr_3$ PNCs as electron donors and $TiO_2$ nanostructures as electron acceptors, achieved by substituting the long-chain oleylamine ligand with a shorter-chain butylamine on the PNC surface post-synthetically. We chose $TiO_2$ as an electron acceptor owing to its type-II band alignment with metal halide perovskites.[20] $TiO_2$ has been extensively studied as thermally and chemically stable, low cost, electron transport layer with suitable band edge positions for perovskite solar cells as well as active photocatalyst for chemical conversions and water splitting.[1b,f,21] We explore aspects such as PL quenching, electron-transfer dynamics, and trap-state modifications resulting from the donor-acceptor interaction. In solution, only butylamine (ButAm)-capped $CsPbBr_3$ PNCs demonstrate interactions with $TiO_2$, resulting in noticeable PL quenching. Conversely, in film states, both ButAm-capped and oleylamine (OlAm)-capped PNCs exhibit robust donor-acceptor interactions with $TiO_2$. The interaction is significantly amplified in $CsPbBr_3$ PNCs with the shorter-chain butylamine ligands, emphasizing efficient electron transfer to $TiO_2$. This study presents pivotal insights into fine-tuning the PNC-$TiO_2$ donor-acceptor interactions through post-synthesis ligand exchange. Further exploration of this approach holds promise in engineering metal halide perovskite-based heterojunctions with efficient charge transfer, benefiting perovskite solar cells, LEDs, and photocatalysis.

**EXPERIMENTAL SECTION**

**Materials**

All the materials were ordered from Sigma Aldrich unless otherwise stated and used without any further purification, which includes: Lead (II) bromide ($PbBr_2$, 98+%), cesium acetate ($CH_3COOCs$, >98%), oleic acid (90%), oleylamine (80-90%), butylamine (99+%), octadecene (90%), toluene (99.5%). Titanium oxide ($TiO_2$) nanoparticle powder was obtained from Kemira Global, Finland.

**Synthesis of CsPbBr$_3$ PNCs**

CsPbBr$_3$ PNCs were synthesized using a hot injection method reported previously with slight modifications.[5a] During the synthesis, $CH_3COOCs$ (0.4 mmol, 77 mg), and oleic acid (1.58 mmol, 500 µL) were added to octadecene (1 mL) in a three-neck flask, and the mixture was heated to 120 °C under an inert atmosphere of argon. After the temperature was stabilized, the mixture was switched to the vacuum condition for 1 h for drying. In parallel, $PbBr_2$ (0.4 mmol, 146 mg), oleic acid (4 mmol, 1.262 mL), and oleylamine (4 mmol, 1.316 mL) were added to octadecene (25 mL) in a separate three-neck flask. The mixture was heated to 120 °C under an argon atmosphere and after the temperature was stabilized, the mixture was switched to the vacuum condition for 1 h for drying. Once the precursors were completely dissolved in octadecene, a clear solution was obtained. Afterward, the temperature of $PbBr_2$ precursor solution was increased to 155 °C, and 1 mL of cesium precursor was injected into it, resulting in the formation of a green precipitate of CsPbBr$_3$ PNCs. The reaction was quenched by placing the reaction flask in an ice-water bath after 2 seconds of the reaction time. The crude CsPbBr$_3$ PNCs were collected by centrifugation at 10,000 rpm for 10 min, and the precipitate was suspended in toluene. The resulting colloidal solution was centrifuged again at 6000 rpm for 10 min to collect the final CsPbBr$_3$ PNC precipitate which was then re-suspended in 6 mL of toluene for further studies.

**Ligand exchange in CsPbBr$_3$ PNCs**

A post-synthesis ligand exchange was performed under ambient conditions to exchange the oleylamine ligand with butylamine ligand on the surface of CsPbBr$_3$ PNCs. A 50 μL of OlAm-capped CsPbBr$_3$ PNCs (6 μmol) in toluene was diluted to 3 mL. Also, 2 μL of butylamine (20 μmol) were added to a 500 μL toluene. The butylamine solution was then added to the CsPbBr$_3$ PNC solution and the mixture was stirred at 500 rpm for 10 minutes under air atmosphere and at room temperature. This was followed by washing of excess ligands using acetonitrile at 5000 rpm for 5 minutes. The obtained precipitate was re-suspended in toluene.

**Growth of TiO$_2$ Nanotube Arrays**

TiO$_2$ nanotube arrays were produced by anodic oxidation of commercial purity, Grade 2 ASTM, titanium sheets that were 0.5 mm thick. From the sheets, 1.5×1.5 cm$^2$ size slabs were cut and polished with P600 SiC paper. The TiO$_2$ slabs were then subjected to sonication in ethanol to remove surface contaminations and anodized in an ethylene glycol solution containing 0.2 M of NH$_4$F and 2 M of distilled water. A cell voltage of 45 V was reached by a linear sweep in 2 minutes and maintained for 30 min in the potentiostatic conditions, as reported in previous studies.[22] At the end of the anodizing procedure, TiO$_2$ slabs were carefully rinsed with distilled water and subjected to annealing at 500 °C for 2 h to obtain oxide crystallization in the anatase phase.

**Preparation of thin film samples**

Thin films of OlAm-capped and ButAm-capped CsPbBr$_3$ PNCs with or without TiO$_2$ nanoparticle powder were spin-coated onto 18 mm × 18 mm glass coverslips (Menzel-Glaser) at 1000 rpm for 2 seconds. Before deposition, the glass coverslips were treated ultrasonically with isopropyl alcohol and acetone respectively, followed by drying using an argon gun. To avoid and minimize the effect of aging and degradation, freshly synthesized PNCs were used, and the measurements were done immediately after the ligand exchange to ensure comparability.

Similarly, freshly-prepared OlAm-capped or ButAm-capped CsPbBr$_3$ PNCs were deposited onto the anodically-grown TiO$_2$ nanotube array by spin coating the colloidal solutions of PNCs at 1000 rpm for 2 seconds, and the measurements were carried out after drying under the argon atmosphere.

**Characterizations**

Fourier transform infrared (FTIR) scans were run by using Spectrum Two FT-IR Spectrometer from Perkin Elmer. The measurements were performed by drying the PNC colloidal solutions on a diamond-attenuated total reflection (ATR) unit in a range from 500 to 4000 cm$^{-1}$.

Scanning Transmission Electron Microscopy (STEM) images of CsPbBr$_3$ PNCs were acquired on a Jeol ARM200CF NeoARM electron microscope at 200 kV acceleration voltage. The samples were prepared by drop-casting the colloidal solution on the carbon-coated TEM grids. Scanning electron microscopy (SEM) images of anodically-grown TiO$_2$ nanotubes with CsPbBr$_3$ PNC deposition were recorded on a Gemini Supra 25 field-emission SEM from Zeiss at an acceleration voltage of 10 kV. Similarly, the SEM images of bare anodically-grown TiO$_2$ nanotubes were recorded on a Gemini SEM from Zeiss at an acceleration voltage of 5 kV. Powder X-ray diffraction (XRD) of the samples was measured using an Aeris X-ray diffractometer from Malvern Panalytical (Cu-k$\alpha$1, 1.5406 Å). The samples were prepared by drop-casting the PNC colloidal solution on a low-background silica disc.

Ultravoilet-visible (UV-Vis) absorption spectra of CsPbBr$_3$ PNCs and TiO$_2$ NPs were recorded in a quartz cuvette using a Lambda 1050+ UV-Vis-NIR spectrometer from Perkin Elmer. Steady-state and time-resolved PL were recorded on Spectrofluorometer FS5 from Edinburg Instrument. For the time-resolved PL measurements, a picosecond laser with 375 nm excitation wavelength was used and the data acquisition was made using the technique of time-correlated single photon count (TCSPC). The decay profiles were tail-fitted with a tri-exponential function $R(t)=A_1 \exp(-t/\tau_1)+A_2 \exp(-t/\tau_2)+A_3 \exp(-t/\tau_3)$ and the intensity-weighted average PL lifetime was calculated using the formula, $\tau_{av}=\frac{A_1\tau_1^2+A_2\tau_2^2+A_3\tau_3^2}{A_1\tau_1+A_2\tau_2+A_3\tau_3}$. Also, the fractional contribution of each decay time to the steady-state intensity was calculated using the formula, $f_i=\frac{A_i\tau_i}{A_1\tau_1+A_2\tau_2+A_3\tau_3}$.

Photoluminescence quantum yields (PL QYs) of the CsPbBr$_3$ PNCs were measured using an absolute method by directly exciting the PNC solution as the sample and the toluene as the reference in an SC-30 integrating sphere module fitted to a Spectrofluorometer FS5 from Edinburg Instrument. During the measurement, the excitation slit was set to 3 nm, and the emission slit was adjusted to obtain a signal level of $1\times10^6$ cps. A wavelength step size of 0.1 nm and an integration time of 0.2 s were used. The calculation of absolute PL QY is based on the formula, PL QY= $\frac{E_{sample}-E_{ref.}}{S_{ref.}-S_{sample}}$, where $E_{sample}$ and $E_{ref}$ are the integrals at the emission region for the sample and the reference, respectively, and $S_{sample}$ and $S_{ref}$ are the integrals at the excitation scatter region for the sample and the reference, respectively. The selection and calculation of integrals from the emission and excitation scattering region and the calculation of absolute PL QY were performed using the FLUORACLE software from the Edinburg Instrument. For steady-state PL spectra and absolute PL QY measurements, the samples were excited at 450 nm.

Fluorescence lifetime imaging microscopy (FLIM) measurements were performed in the confocal configuration in an ambient atmosphere on a MicroTime200 fluorescence microscope from PicoQuant equipped with 440 nm picosecond laser, 60× objective, PMA Hybrid single photon detector, and PicoHarp 300 TCSPC module. The excitation spot size was estimated at 550 nm (FWHM).

## RESULTS AND DISCUSSION

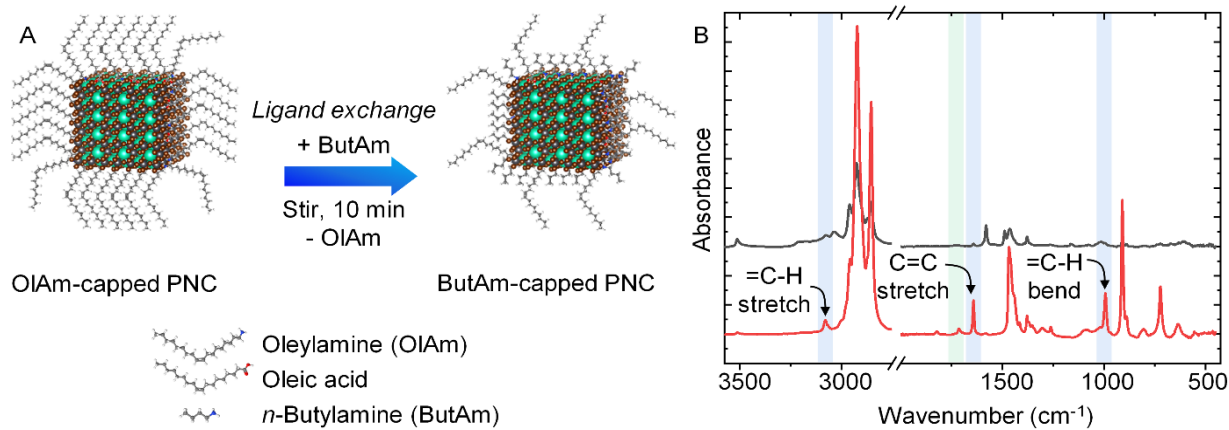

**Figure 1.** (A) Scheme showing the exchange of long-chain oleylamine ligands on the surface of PNCs with short-chain butylamine ligands. (B) FTIR spectra of CsPbBr$_3$ PNCs before (red) and after (black) ligand exchange. The light blue regions correspond to the vibration frequencies of oleylamine, and the light green region corresponds to the vibration frequency of oleic acid.

We investigated how altering the chain length of aliphatic amine ligands impacts the CsPbBr$_3$ PNC-TiO$_2$ donor-acceptor interactions in both solution and film states. We employed a post-synthesis ligand exchange approach to functionalize the PNC surface with short-chain butylamine ligands. Initially, we synthesized CsPbBr$_3$ PNCs using a hot-injection method reported earlier by Protessescu et al. with slight modifications.[5a] Oleic acid and oleylamine were used as capping ligands for the PNCs. The as-synthesized PNCs were directly subjected to ligand exchange without any additional washing steps. Figure 1 A shows the scheme for post-synthesis ligand exchange on as-synthesized CsPbBr$_3$ PNCs. The ligand exchange was conducted under ambient conditions by introducing butylamine to the OlAm-capped PNCs in toluene while stirring at room temperature. We optimized the amount of butylamine to 2 µL (20 µmol) per 3 mL (6 µmol) of PNC solution during the ligand exchange without causing damage to the original OlAm-capped PNCs. Indeed, we observed that adding a significant amount of butylamine (≥10 µL) to the same amount of OlAm-capped CsPbBr$_3$ PNCs causes a noticeable change in the color of the colloidal solution, shifting from green to colorless and quenching of PL (Supporting Information Figure S1). Previous studies have highlighted a similar effect concerning the ligand-assisted degradation of PNCs or their transformation to other phases.[5b,23] After the ligand exchange, the sample was washed thoroughly by adding acetonitrile to remove excess ligands, and the resulting ButAm-capped PNCs were collected by centrifugation and redispersed in toluene for further studies. We used Fourier transform infrared (FTIR) spectroscopy in attenuated total reflection mode to confirm the ligand exchange process. Figure 1 B provides a comparison of the FTIR spectra of OlAm-capped $CsPbBr_3$ PNCs with those of ButAm-capped PNCs after the ligand exchange. Following the ligand exchange, IR absorption bands at approximately 3080 $cm^{-1}$, 1642 $cm^{-1}$, and 990 $cm^{-1}$ that are related to =C-H stretching, C=C stretching, and =C-H bending vibrations, respectively,[24] and are closely associated with oleylamine or oleic acid showed a significant reduction or disappearance in ButAm-capped PNCs. Notably, the IR band at approximately 1715 $cm^{-1}$, associated with the C=O stretching vibration of an organic acid,[24] also diminished in ButAm-capped PNCs compared to the OlAm-capped ones. These findings indicate that during the ligand exchange, butylamine largely replaces oleylamine on the surface of $CsPbBr_3$ PNCs. Concurrently, the concentration of oleic acid ligands on the PNC surface also decreases. These results align with the dynamic ligand binding mode, where oleate binds to the PNC surface as an ion pair with oleylammonium.[18a] Hence, replacing oleylamine with butylamine on the PNC surface results in the removal of oleic acid, reducing its concentration on the PNC surface.

We examined the morphology and crystal structure of OlAm- and ButAm-capped $CsPbBr_3$ PNCs using high-resolution scanning transmission electron microscopy (HR-STEM) and powder X-ray diffraction (XRD), as detailed in Figure 2. The STEM images show that the OlAm-capped PNCs are cube-shaped with an average size of 12 nm (Figure 2 A, Supporting

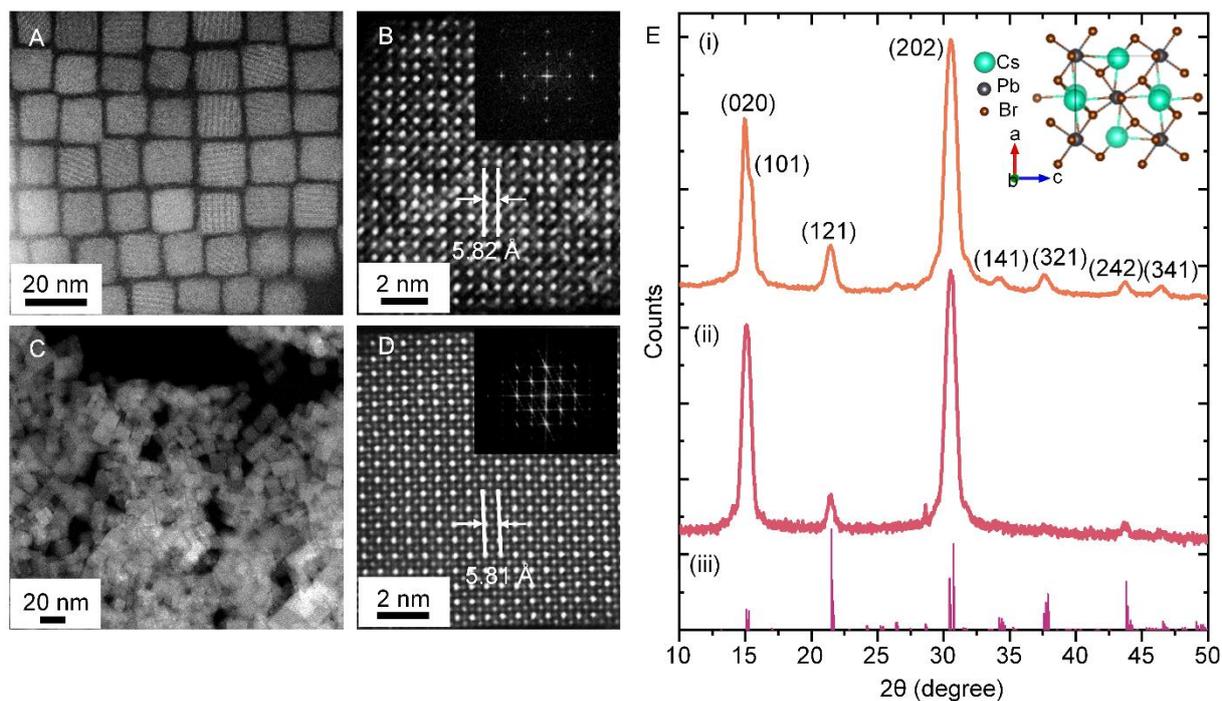

**Figure 2.** (A, C) STEM and (B, D) HR-STEM images of as-synthesized OlAm-capped (A, B), and ligand-exchanged ButAm-capped (C, D) $CsPbBr_3$ PNCs. The insets in B and D are the corresponding Fast-Fourier Transformed (FFT) images. (E) Powder XRD patterns of (i) ButAm-capped and (ii) OlAm-capped $CsPbBr_3$ PNCs which are compared with (iii) ref. 26. The inset shows the crystal structure of orthorhombic $CsPbBr_3$.

Information Figure S2 A) which is maintained after ligand exchange with butylamine (Figure 2 C, Supporting Information Figure S2 B). Conversely, the ButAm-capped $CsPbBr_3$ PNCs exhibited aggregation on the TEM grid due to reduced interparticle distance caused by the shorter chain ligands on the PNC surface. Further, the HR-STEM image of a ButAm-capped $CsPbBr_3$ PNC in Figure 2 D displays a lattice spacing of 5.81 Å, consistent with that of an OlAm-capped PNC (Figure 2 B) and corresponds to the (101) plane of an orthorhombic $CsPbBr_3$ crystal structure.[25] This is further supported by the powder XRD patterns of OlAm- and ButAm-capped $CsPbBr_3$ PNCs shown in Figure 2 E. In the XRD pattern, the diffraction peak at a 2θ angle of 15° in OlAm-capped $CsPbBr_3$ PNC (Figure 2 E i) split into (020) and (101) peaks, corresponding to the orthorhombic crystal phase.[26] Nevertheless, the broadening of XRD peaks hinders the clear observation of peak splitting for both samples when compared with the bulk CsPbBr$_3$, which is attributed to the small crystal size.[27] Still, all the diffraction peaks in both cases align well with those of the orthorhombic CsPbBr$_3$ reported in the literature.[26]

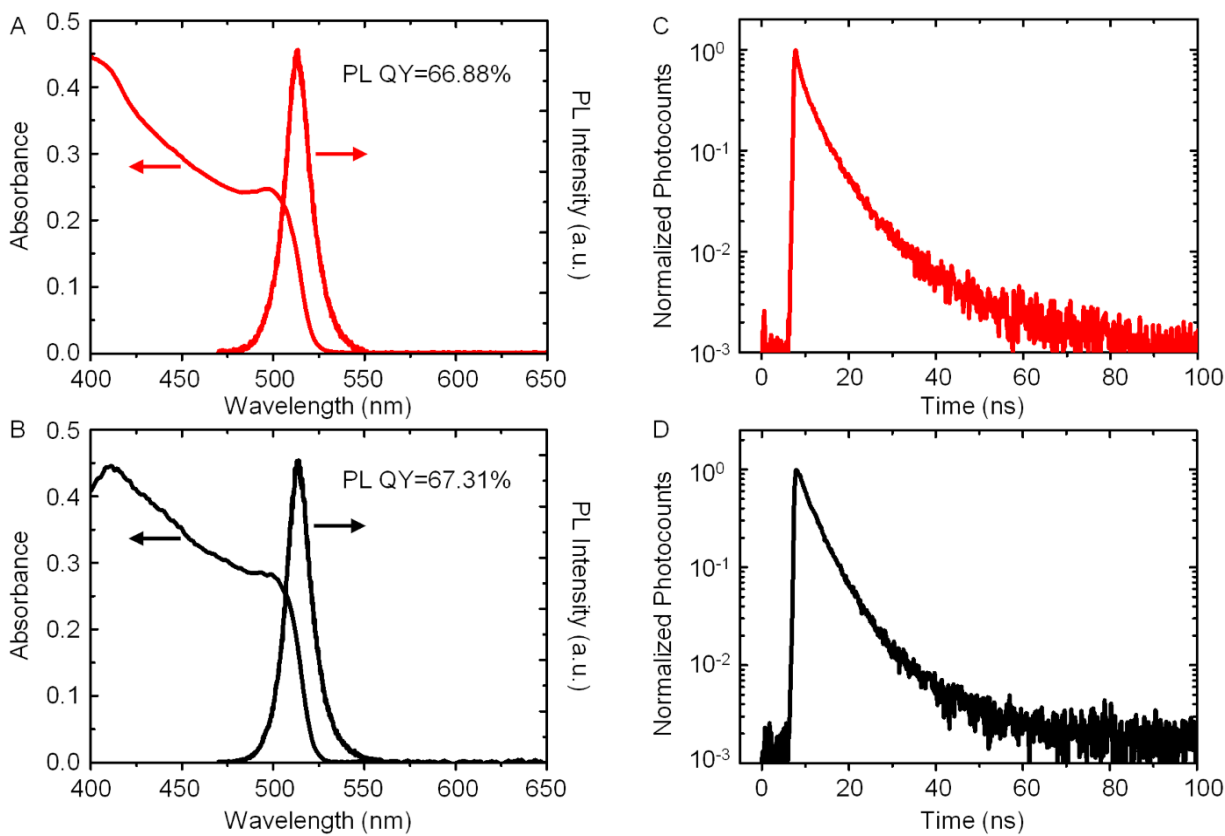

**Figure 3.** (A, B) Absorption and PL spectra of OlAm-capped (A) and ButAm-capped (B) CsPbBr$_3$ PNCs. (C, D) PL decay profiles of OlAm-capped (A) and ButAm-capped (B) CsPbBr$_3$ PNCs.

We characterized the optical and PL properties of CsPbBr$_3$ PNCs in solution before and after the ligand exchange using absorption, steady-state, and time-resolved PL spectroscopy. Figures 3 A and B show absorption and PL spectra for OlAm- and ButAm-capped PNCs, respectively, showcasing comparable results. The spectra display the absorption onset at 528 nm with the near band-edge exciton absorption occurring at 500 nm. The PL peak is observed at 510 nm, featuring a full width at half maximum of 18 nm. Further, both the samples exhibit bright green emission with an absolute PL QY of 67 % (Supporting Information Figure S3). To understand the exciton recombination dynamics, we recorded the PL decay profiles for OlAm- and ButAm-capped $CsPbBr_3$ PNC solutions using a time-correlated single-photon counting (TCSPC) system, which are shown in Figures 3 C and D, respectively. We analyzed the decays by fitting them with a tri-exponential decay function and calculated the intensity-weighted average PL lifetimes ($\tau$). The obtained $\tau$ values are 5.96 ns for OlAm-capped PNCs and 5.50 ns for ButAm-capped PNCs, demonstrating similar PL lifetime in both cases. These findings affirm that the $CsPbBr_3$ PNCs maintain their optical and PL properties after undergoing post-synthesis ligand exchange.

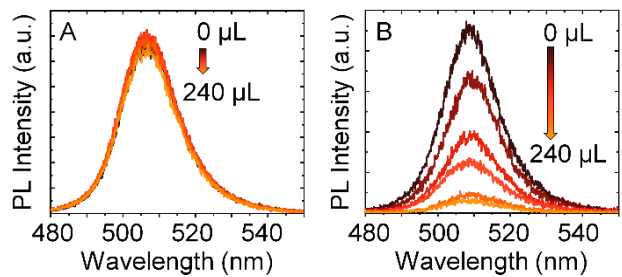

**Figure 4.** (A, B) PL spectra showing PL quenching of $CsPbBr_3$ PNCs by $TiO_2$ in solution before (A) and after (B) ligand exchange.

With successful post-synthesis ligand exchange that maintains the morphology, crystal structure, optical absorption and PL properties of $CsPbBr_3$ PNCs, we investigated the influence of ligand chain length on the donor-acceptor interactions between PNCs and $TiO_2$ nanoparticles (NPs) and nanotubes (NTs) in the solution and film state. In the solution phase, we began by examining the PL spectra of OlAm- and ButAm-capped PNCs with systematic addition of different volumes of $TiO_2$ suspension in toluene. The colloidal PNC solutions for both samples were suitably diluted to an optical density of 0.5 to ensure a comparable nanocrystal concentration and mitigate any inner filtering effects. To these solutions, we added $TiO_2$ NP (size 50 nm to 150 nm, Supporting Information Figure S4 A) suspension at different volumes increasing from 0 to 240 µL, corresponding to final concentrations ranging from 0 to 120 mM. The stock suspension was prepared by mixing 1 g of $TiO_2$ NPs in 14 mL of toluene. The structural and optical characterization of $TiO_2$ NPs are provided in the Supporting Information

Figure S5. As shown in Figure 4 A, we observed constant PL intensities for the OlAm-capped CsPbBr$_3$ PNC solution, displaying no PL quenching in presence of TiO$_2$. However, in the case of the ButAm-capped PNC solution, the PL intensities decreased (Figure 4 B) with an increase in the added volume of TiO$_2$ suspension. Here, the PL spectra obtained after each addition of TiO$_2$ suspension in toluene is corrected for the dilution effect, the details of which are provided in the Supporting Information. Also, the PNC solutions were colloidally stable at different dilution as indicated by the consistent absorption spectra of CsPbBr$_3$ PNCs at different volume of with TiO$_2$ suspension added as shown in Supporting Information Figure S6. These results suggest that the short-chain ligands enhance the photophysical interactions between the charge donor and acceptor species. Considering the type-II band alignment between CsPbBr$_3$ and TiO$_2$,[20] we anticipate that the donor-acceptor interactions in our case involve electron transfer. Nevertheless, to understand this further, one has to look closer into the dynamic aspects of the quenching interaction.

To comprehend the mechanism of PL quenching in solution, we plotted Stern-Volmer plots[28] for the OlAm- and ButAm-capped CsPbBr$_3$ PNCs at different concentrations of TiO$_2$ as a quencher by taking the PL intensity values from Figure 4. The results are shown in Figure 5 A. The Stern-Volmer plot for OlAm-capped PNCs exhibits a flat line (zero slope), confirming no PL quenching effect from TiO$_2$. Conversely, the Stern-Volmer plot for ButAm-capped CsPbBr$_3$ PNC solution shows a nearly linear trend at lower concentrations, curving towards the y-axis at higher concentrations. This behavior of the Stern-Volmer plot suggests a combination of static and dynamic PL quenching.[28] In static PL quenching, the quencher forms a non-luminescent complex with the light-emitting species before it is excited. In dynamic PL quenching, the quencher collides with the light-emitting species in its excited state, preventing it from emitting light. In both cases, the Stern-Volmer plot yields a straight line with a slope. The distinguishing

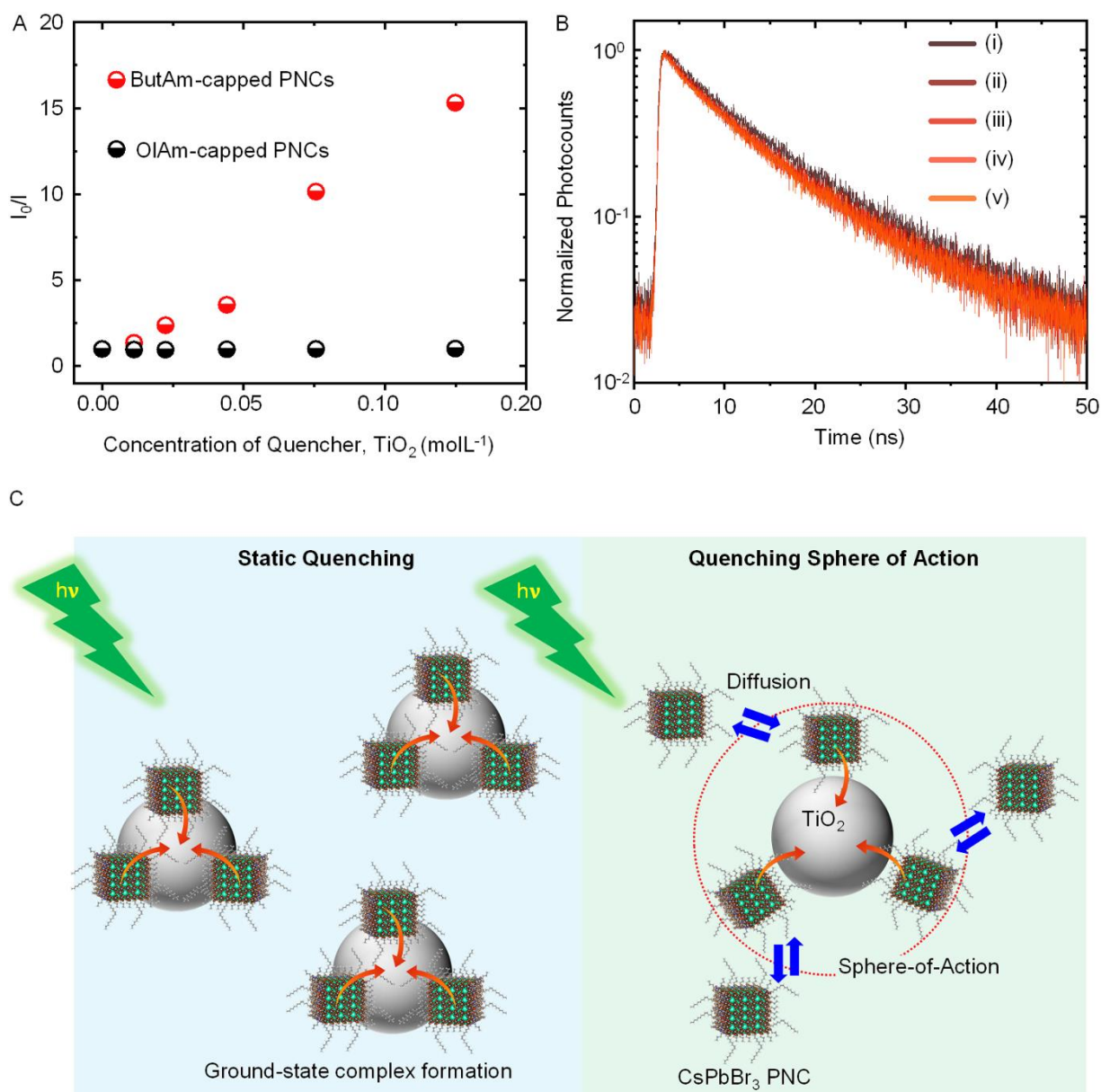

**Figure 5.** (A) Stern-Volmer plots for OlAm-capped and ButAm-capped $CsPbBr_3$ PNC colloidal solutions in the presence of $TiO_2$ as quencher. (B) PL decay profiles of ButAm-capped $CsPbBr_3$ PNC colloidal solution obtained after adding (i) 0 µL, (ii) 20 µL, (iii) 40 µL, (iv) 80 µL, and (vi) 240 µL of $TiO_2$ NP suspension in toluene. (C) Mechanism of PL quenching of $CsPbBr_3$ PNCs by $TiO_2$ NPs in solution.

factor between these PL quenching mechanisms is the PL lifetimes of the light-emitting species at varying quencher concentrations. For static quenching, PL lifetime remains constant regardless of quencher concentration. Conversely, in dynamic quenching, PL lifetime decreases as quencher concentration increases.[28] In our study, the PL lifetime of ButAm-capped $CsPbBr_3$

PNCs remained constant across all $TiO_2$ concentrations, as depicted in Figure 5 B. This contradicts the non-linear behavior observed in the Stern-Volmer plot in our case if we assume a prevalence of mixed static and dynamic quenching. In simple terms, we expected the PL lifetime to decrease as the quenching shifted from static to dynamic, but this was not observed in the PL decays. To better understand this, we looked beyond the traditional Stern-Volmer plot and considered deviations, incorporating mechanisms involving static quenching and quenching by the sphere of action,[28,29] as shown in Figure 5 C.

During the static quenching, we assume a ground-state complex formation by the adsorption of ButAm-capped $CsPbBr_3$ PNCs onto the surface of $TiO_2$ (Figure 5 C). The short-chain butylamine ligand and the reduced concentration of oleic acid ligand on the surface of PNCs facilitated this adsorption. For the upward bending of the Stern-Volmer plot at the higher concentrations of $TiO_2$, we propose PL quenching by the sphere of action mechanism in addition to the static quenching. Figure 5 C shows the quenching by the sphere of action mechanism where $CsPbBr_3$ PNC and $TiO_2$ do not form a stable ground-state complex. Instead, an apparent static quenching results within a volume where the PNC emitter and $TiO_2$ quencher are very close to each other at the time of excitation. Within this sphere of action, there exists a high probability that the PL quenching occurs before the PNC emitter and the $TiO_2$ quencher diffuse apart. Further, the probability of PL quenching increases with increasing the concentration of the quencher. On the other hand, the presence of long-chain oleylamine ligands and a higher concentration of oleic acid on the surface of as-synthesized $CsPbBr_3$ PNCs in conjunction with solvent intervention prevented their interaction with $TiO_2$, resulting in no static or dynamic PL quenching. In line with our findings, recently, Vinçon et al. reported the interaction of metal salt $BiBr_3$ with lecithin-capped $CsPbBr_3$ QDs through the sphere of action model which quenches PL in QDs but without significantly changing the PL lifetime.[29] While they attribute the PL quenching of $CsPbBr_3$ QDs by the controlled addition of Bi defects to the QD surface, we discuss PL quenching of ButAm-capped $CsPbBr_3$ PNCs by $TiO_2$ NPs in solution in terms of donor-acceptor interactions leading to the electron transfer. Further, it is important to consider the artifact caused by the inner filtering effect which may result in PL quenching.[30] Here, at a fairly high concentration of quencher that is added to the emitter in solution, the excitation light might get blocked by scattering or absorption by the quencher before reaching the emitter. This can also result in a lowering of PL intensity. In our case, such an inner filtering effect is trivial since we do not observe any decrease in PL intensity by adding the same concentration/amount of $TiO_2$ to the OlAm-capped $CsPbBr_3$ PNC solution as in the case of ButAm-capped ones. The exclusion of inner filtering effect is further supported by the absorption spectra in Supporting Information Figure S6, whose features are consistent over the range of $TiO_2$ suspension added to PNC solution.

In solution, we observed ligand chain length-dependent PL quenching which is indicative of potential donor-acceptor interactions that could facilitate charge transfer between $CsPbBr_3$ PNCs and $TiO_2$ NPs. ButAm-capped $CsPbBr_3$ PNCs showed enhanced PL quenching and hence efficient electron transfer to $TiO_2$ as compared with OlAm-capped PNCs when photoexcited. However, the unchanged PL decay profiles regardless of the $TiO_2$ NP concentration in the solution limited us from discussing the influence of donor-acceptor interactions on exciton recombination dynamics. Therefore, we investigated the $CsPbBr_3$-$TiO_2$ donor-acceptor system in its solid state with different chain length ligands on the PNC surface using a confocal fluorescence lifetime imaging microscopy (FLIM) or a time-resolved PL spectroscopy. Further, to unveal the influence of $TiO_2$ microstructure on electron transfer in solid state we studied two systems, one involved a film formed by depositing a mixture of $TiO_2$ NPs and OlAm-capped or ButAm-capped $CsPbBr_3$ PNCs from the solution on a glass slide and the other was OlAm-capped or ButAm-capped $CsPbBr_3$ PNC layers deposited on the top of an anodically-grown $TiO_2$ NT array on a $TiO_2$ substrate. The deposition in either of the cases was performed by spin coating followed by drying under argon. Contrary to the solution state, the samples in the solid state, regardless of ligand chain length, showed rapid PL decay in the presence of TiO$_2$, indicating an accelerated exciton recombination due to enhanced donor-acceptor interaction.[31] The formation of a donor-acceptor interface between CsPbBr$_3$ PNCs and TiO$_2$ NPs in solid state is observed on a TEM grid, which is shown in the Supporting Information Figure S4 B. Previously, we demonstrated PL quenching revealed by the decrease in PL lifetimes across the heterojunction film formed by OlAm-capped FAPbBr$_3$ or CsPbBr$_3$ PNCs with TiO$_2$ and attributed the effect to carrier diffusion-controlled electron transfer from PNCs to TiO$_2$ at the interface.[32] However, our previous study did not shed light on the influence of surface ligands on electron transfer in the solution or the film state. While, in this study, we show ligand chain length-dependent donor-acceptor interactions between CsPbBr$_3$ PNCs and TiO$_2$ in solution and films and discuss different mechanisms involved therein.

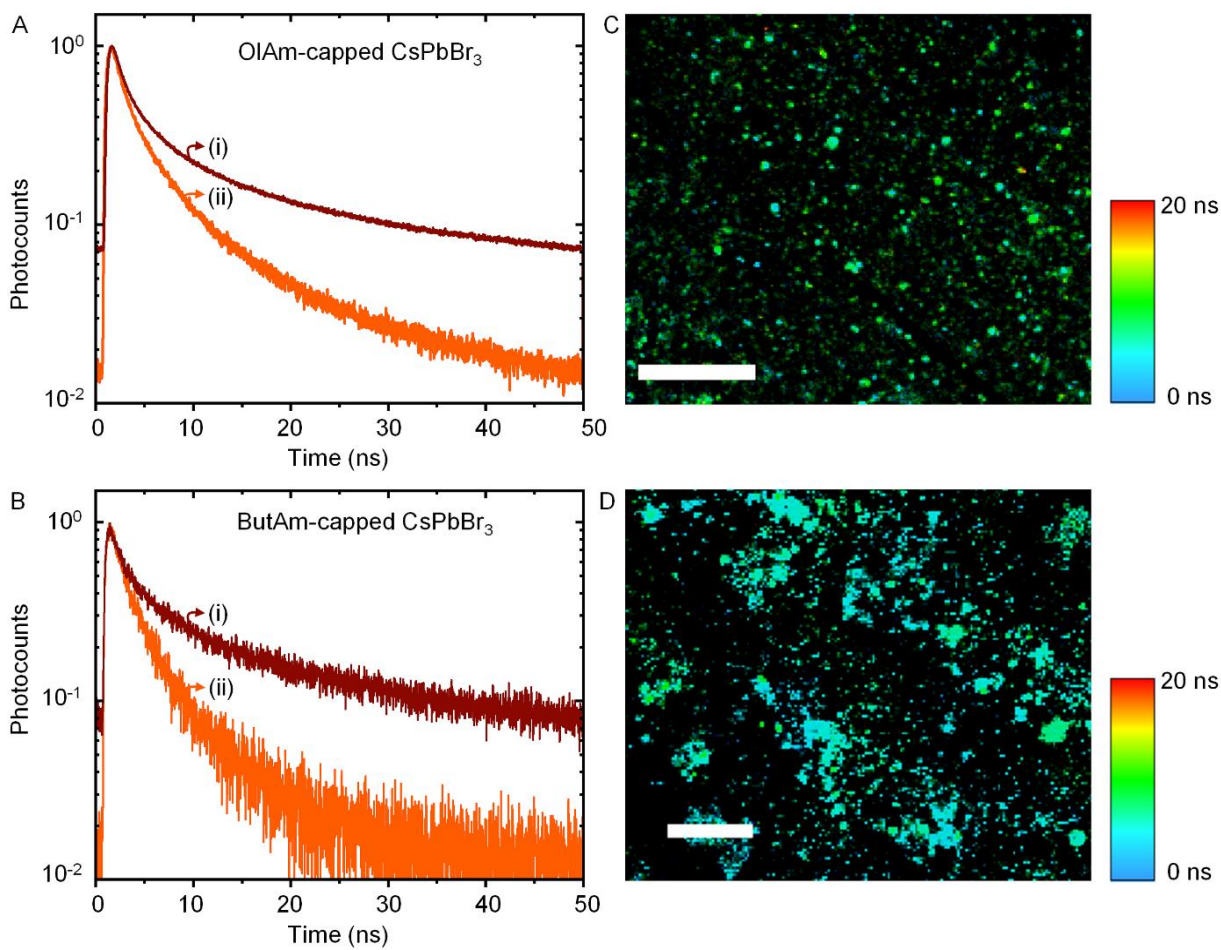

**Figure 6.** (A, B) PL decay profiles of OlAm-capped (A) and ButAm-capped (B) CsPbBr$_3$ PNC films before (i) and after (ii) mixing PNCs with TiO$_2$. (C, D) FLIM images of TiO$_2$ mixed OlAm-capped (C) and ButAm-capped (D) CsPbBr$_3$ films. The scale bars are 25 µm and 100 µm in C and D, respectively.

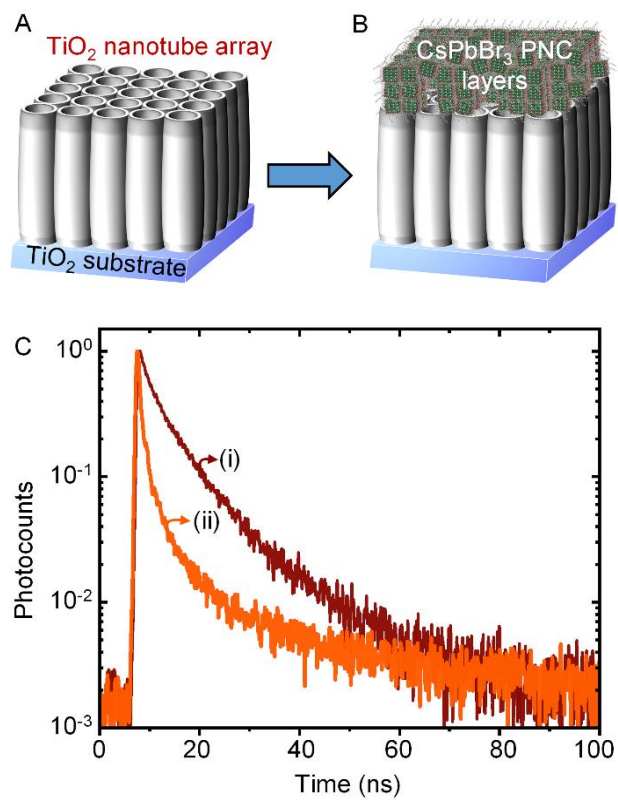

**Figure 7.** (A, B) Scheme showing anodically grown TiO$_2$ NT array on a TiO$_2$ substrate (A) without CsPbBr$_3$ PNCs and (B) with CsPbBr$_3$ PNCs deposited on the top. (C) PL decay profiles of ButAm-capped CsPbBr$_3$ PNCs deposited on (i) a glass coverslip and (ii) the top of the TiO$_2$ NT array.

The FLIM images for OlAm-capped and ButAm-capped CsPbBr$_3$ PNCs mixed with TiO$_2$ NPs are shown in Figures 6 C and D, respectively and the corresponding decay profiles are presented in Figures 6 A and B. The OlAm-capped and ButAm-capped CsPbBr$_3$ PNC films without TiO$_2$ show comparable average PL lifetime values of about 14 ns. Interestingly, when PNCs were mixed with TiO$_2$ NPs to form films, the average PL lifetime decreased, especially for ButAm-capped CsPbBr$_3$ PNCs (ca. 6 ns), compared to OlAm-capped PNCs (ca. 9 ns). This decrease suggested that the donor-acceptor interactions occurred in the film state for both OlAm-capped and ButAm-capped CsPbBr$_3$ PNCs, and this interaction probably led to an enhanced electron transfer for PNCs with short-chain butylamine ligands. Furthermore, the contradicting results for PL quenching that occurs in solid film but not in solution for OlAm-capped PNCs suggest that, in the solution state, not only the ligand chain length but also the solvent interferes with the donor-acceptor interactions.

Figures 7 A and B are schematic representations of the TiO$_2$ NT array before and after the CsPbBr$_3$ PNC deposition. The corresponding scanning electron microscopic (SEM) images are provided in the Supporting Information Figure S7. Additionally, Figure 7 C shows the PL decay profiles for ButAm-capped CsPbBr$_3$ PNCs deposited on the glass slides and top of the TiO$_2$ NT array. The PL decay is rapid for PNCs deposited on top of the TiO$_2$ NT array compared to those on a bare glass substrate, aligning with the trend discussed earlier. To further analyze this, we compared the PL lifetime components ($\tau_i$), corresponding fractional contribution ($f_i$) of each decay time to the steady-state intensity, and the average $\tau$, obtained by fitting the PL decays with tri-exponential decay function, for these two systems: CsPbBr$_3$ PNC-TiO$_2$ NP film and CsPbBr$_3$ PNC-TiO$_2$ NT arrays. This comparison also included CsPbBr$_3$ PNC films without TiO$_2$. The results are summarized in Table 1. For PNC films without TiO$_2$, the average PL lifetime is largely contributed by the longest component $\tau_3$ which is 19 ns (66%) for OlAm-capped CsPbBr$_3$ PNCs and 18 ns (72%) for ButAm-capped CsPbBr$_3$ PNCs. On the other hand, the shortest component $\tau_1$, which is 1.18 ns for OlAm-capped CsPbBr$_3$ PNCs and 1.10 for ButAm-capped ones, contributes less than 10% to the average PL lifetime. Additionally, the intermediate component $\tau_2$ (4.2 ns for OlAm-capped and 3.8 ns for ButAm-capped CsPbBr$_3$ PNCs) contributes more than 20% to the average PL lifetime. Earlier reports suggest that $\tau_1$ is associated with the trapping of charge carriers in defect-related shallow or deep trap states.[33] Charge carriers quickly transition from the band edge to these trap states, depleting the concentration of carriers at the band edge and causing nonradiative recombination of excitons

due to minimal detrapping.[34] On the other hand, $\tau_2$ is influenced by exciton recombination at the band edge involving both radiative and nonradiative processes, as well as charge carrier trapping in defect-related trap states. At this stage, the shallow traps are not fully saturated with charge carriers, affecting the trapping and detrapping rates. Despite a reduced trapping rate, the detrapping rate remains inefficient, resulting in a PL lifetime only slightly longer than $\tau_1$.[34] The longest component $\tau_3$ correlates with radiative recombination at the band edge, particularly when shallow trap states are nearly saturated with charge carriers. This leads to a momentary trapping and subsequent detrapping of charge carriers, slowing down the radiative recombination of excitons at the band edge.[34,35] The prevalence of $\tau_3$ in both film samples without $TiO_2$ NPs indicates that excitons primarily recombine radiatively, consistent with the high absolute PL QY values observed for these samples.

**Table 1.** PL lifetime components ($\tau_i$), corresponding fractional contribution ($f_i$) of each decay time to the steady-state intensity, and intensity-weighted average PL lifetimes ($\tau$) obtained from

|  | $\tau_1$ (ns) | $f_1$ | $\tau_2$ (ns) | $f_2$ | $\tau_3$ (ns) | $f_3$ | Average $\tau$ (ns) |
|---|---|---|---|---|---|---|---|
| OlAm-$CsPbBr_3$ PNC film | 1.18 | 0.099 | 4.20 | 0.241 | 19.00 | 0.660 | 13.67 |
| ButAm-$CsPbBr_3$ PNC film | 1.10 | 0.062 | 3.80 | 0.215 | 18.00 | 0.723 | 13.90 |
| OlAm-$CsPbBr_3$ PNC+$TiO_2$ NP film | 1.50 | 0.232 | 5.10 | 0.435 | 19.00 | 0.333 | 8.89 |
| ButAm-$CsPbBr_3$ PNC+$TiO_2$ NP film | 1.40 | 0.221 | 3.60 | 0.454 | 12.00 | 0.325 | 5.84 |
| ButAm-$CsPbBr_3$ PNCs on $TiO_2$ NT array | 0.35 | 0.230 | 1.83 | 0.434 | 9.33 | 0.332 | 3.97 |

multiexponential fitting of PL decay profiles of Olm-capped and ButAm-capped CsPbBr$_3$ PNC film, CsPbBr$_3$-TiO$_2$ film and CsPbBr$_3$ PNCs deposited on the top of a TiO$_2$ NT array.

We observed a distinct shift in the contribution of the radiative and nonradiative exciton recombination processes on the overall PL lifetime of CsPbBr$_3$ PNCs in their films in the absence or presence of TiO$_2$ (Table 1). Specifically, in the OlAm-capped CsPbBr$_3$ PNC-TiO$_2$ NP film, $\tau_1$, $\tau_2$, and $\tau_3$ values remained unchanged compared to the film without TiO$_2$. However, their relative contributions to the average PL lifetime shifted from the longest component $\tau_3$ (66 to 33%) to the shortest $\tau_1$ (10 to 23%) and intermediate $\tau_2$ (24 to 44%). This suggests that the introduction of TiO$_2$ NPs as electron acceptors does not create fresh nonradiative decay paths for excitons in OlAm-capped CsPbBr$_3$ PNCs within the film. Instead, TiO$_2$ NPs likely favor the nonradiative recombination affecting the overall rate at which excitons recombine, possibly through the modification of trap states. Further study is required to completely understand the influence of charge acceptors on the modification of existing trap states in metal halide perovskites. On the other hand, we observed a shift from the contribution of $\tau_3$ (72 to 33%) to $\tau_1$ (6 to 22%) and $\tau_2$ (22 to 45%) in the average PL lifetime, along with a decrease in $\tau_3$ value from 18 ns to 12 ns in the case of ButAm-capped CsPbBr$_3$ PNC-TiO$_2$ NP film. This implies that TiO$_2$ NPs not only favor the nonradiative exciton recombination in ButAm-capped CsPbBr$_3$ PNCs but also impact the radiative recombination by altering the rate of electron detrapping from the shallow trap states. We propose that in ButAm-capped CsPbBr$_3$ PNCs, electron transfer from shallow trap states to TiO$_2$ NPs competes with the detrapping of electrons to the band edge. The former process takes precedence, resulting in faster radiative recombination of the untrapped excitons at the band edge. In the case of ButAm-capped CsPbBr$_3$ PNCs on the TiO$_2$ NT array, the alteration in the contributions of PL lifetime components $\tau_1$, $\tau_2$, and $\tau_3$ to the average PL lifetime aligns with that seen in PNC-TiO$_2$ NP films (Table 1). However, the values of all three PL lifetime components decrease notably to 0.35 ns for $\tau_1$, 1.83 ns for $\tau_2$, and 9 ns for $\tau_3$. Particularly, the $\tau_1$ decrease is substantial compared to PNC films with $TiO_2$ NPs. These findings highlight that, in the ButAm-capped $CsPbBr_3$ PNC-$TiO_2$ NT array system, $TiO_2$ not only impacts existing recombination processes but also creates new nonradiative decay paths for excitons. This can be attributed to the efficient charge transfer from $CsPbBr_3$ PNCs to $TiO_2$ NTs.

**CONCLUSION**

We explored the impact of altering ligand chain lengths on $CsPbBr_3$ PNCs by employing a post-synthesis ligand exchange strategy. This method allowed us to replace long-chain ligands with shorter ones, enabling the study of distance-dependent photophysical interactions. The length of the ligand chain proved to be a critical factor affecting donor-acceptor interactions, influencing PL quenching and electron transfer within $CsPbBr_3$ PNCs. Our investigation focused on understanding these interactions in the $CsPbBr_3$-$TiO_2$ system, both in solution and film states. In the solution phase, we observed enhanced donor-acceptor interactions for PNCs with short-chain butylamine ligands, leading to a mixed PL quenching interaction encompassing static quenching and quenching through the sphere of action. Conversely, $CsPbBr_3$ PNCs with oleylamine ligands in solution showed no notable changes in PL properties in the presence of $TiO_2$ NPs, suggesting poor donor-acceptor interactions due to the presence of long-chain ligands on the PNC surface. In dried films, both long-chain and short-chain ligand-capped $CsPbBr_3$ PNCs exhibited rapid PL decay in the presence of $TiO_2$, implying accelerated exciton recombination. Intriguingly, solid-state PL quenching for long-chain ligand-capped PNCs contradicted the solution state, highlighting the influence of the solvent on donor-acceptor interactions. Additionally, the exciton recombination dynamics in PNCs are affected more by ligand chain length in the presence of $TiO_2$ NTs compared to $TiO_2$ NPs, illustrating the effect of the $TiO_2$ microstructure on charge transfer. In summary, our study sheds light on the intricate interplay between $TiO_2$ and the ligands of $CsPbBr_3$ PNCs, providing crucial insights for their application in optoelectronics.

**ASSOCIATED CONTENTS**

*Supporting Information*

**ACKNOWLEDGEMENTS**

S. A. B. S. acknowledges the Italian Ministry of University and Research (MUR) and Erasmus+ Doctoral Mobility 2022/2023. S. G. acknowledges Alexander von Humboldt-Stiftung/Foundation for the postdoctoral research fellowship. C. K. acknowledges the European Regional Development Fund of the European Union for funding the PL spectrometer (GHS-20-0035/P000376218) and X-ray diffractometer (GHS-20-0036/P000379642) and the Deutsche Forschungsgemeinschaft (DFG) for funding an electron microscope ThermoFisher Talos L120C (INST 264/188-1 FUGG) and for supporting the collaborative research center SFB 1477 "Light-Matter Interactions at Interfaces (LiMatI)", project number 441234705 (W03 and W05). We thank Dr. Kevin Oldenburg, Center for Interdisciplinary Electron Microscopy (ELMI-MV), Department "Life, Light & Matter", University of Rostock, Germany for high-resolution transmission electron microscopy images. We also thank Prof. Dr. Sylvia Speller and M.Sc. Ronja Piehler, Institute of Physics, University of Rostock, Germany for supporting us with the scanning electron microscope.